# A Machine Learning-Driven Website Platform and Browser Extension for Real-Time Risk Scoring and Fraud Detection for Website Legitimacy Verification and Consumer Protection

**Md Kamrul Hasan Chy[1], Obed Nana Buadi[2]**

[1] Department of Computer Information Systems & Analytics, University of Central Arkansas, Conway, USA

[2] Department of Economics and Finance, University of Texas at El Paso, El Paso, USA.

*Abstract*—**This paper introduces a Machine Learning-Driven website Platform and Browser Extension designed to quickly enhance online security by providing real-time risk scoring and fraud detection for website legitimacy verification and consumer protection. The platform works seamlessly in the background to analyze website behavior, network traffic, and user interactions, offering immediate feedback and alerts when potential threats are detected. By integrating this system into a user-friendly browser extension, the platform empowers individuals to navigate the web safely, reducing the risk of engaging with fraudulent websites. Its real-time functionality is crucial in e-commerce and everyday browsing, where quick, actionable insights can prevent financial losses, identity theft, and exposure to malicious sites. This paper explores how this solution offers a practical, fast-acting tool for enhancing online consumer protection, underscoring its potential to play a critical role in safeguarding users and maintaining trust in digital transactions. The platform's focus on speed and efficiency makes it an essential asset for preventing fraud in today's increasingly digital world.**

## I. INTRODUCTION:

In the digital age, e-commerce has become a cornerstone of global business, facilitating billions of transactions daily. However, as online shopping continues to grow, so too does the prevalence of online fraud, posing a significant threat to both consumers and businesses. Fraudulent websites, phishing attacks, and deceptive online practices are increasingly sophisticated, making it harder for consumers to discern legitimate platforms from malicious ones. These fraudulent activities result in substantial financial losses, diminished trust in online transactions, and, in extreme cases, threats to national security.

The rapid advancement of Internet technology has led to a surge in online financial transactions, making them a popular method for purchasing goods and services [1]. However, this widespread adoption of digital payments, particularly credit cards, has also heightened the potential for abuse [1]. Online classified ads, e-commerce platforms, and even dating sites have become breeding grounds for various types of fraud, including deceptive advertisements, commercial extortion, and identity theft [2].

As cyber threats continue to evolve, innovative solutions are essential to protect consumers from fraudulent activities and misinformation. This system leverages advanced machine learning algorithms to analyze vast amounts of data in real-time, enabling quick and accurate risk assessments of websites and online transactions. Similar approaches have been successfully implemented in other domains, such as detecting phishing websites with high accuracy (up to 97.5%) using browser extensions [3]. By incorporating machine learning techniques, the platform can identify complex patterns and relationships that may indicate fraudulent activities, significantly enhancing consumer protection.

The browser extension serves as the user-facing component, seamlessly integrating with popular web browsers to provide immediate feedback on website credibility and potential risks. This approach has been proven effective in other contexts, such as the "Deep Breath" browser extension, which warns users about potentially misleading content online [4]. By offering real-time notifications, users can make informed decisions while browsing the internet or conducting online transactions. One of the key features of this platform is its real-time risk scoring mechanism. By analyzing various parameters such as domain age, SSL certificate validity, user reviews, and historical fraud reports, the system can quickly calculate a risk score for each website or transaction. This approach is similar to other security monitoring extensions that use machine learning to detect irregularities in browser extensions and network requests [5].

Moreover, the platform's fraud detection capabilities go beyond simple website verification. By employing advanced machine learning techniques, including potentially graph neural networks, the system can identify complex patterns and relationships that may indicate fraudulent activities [6]. This multi-faceted approach to fraud detection is crucial in today's online environment, where phishing attacks and other security threats continue to pose significant risks to internet users [7].

By combining real-time risk scoring, fraud detection, and website legitimacy verification in a user-friendly browser extension and a website platform, these platforms represent a significant advancement in online consumer protection. As demonstrated by similar projects in various fields, the integration of machine learning with other technologies can facilitate practical applicability in areas such as food safety analysis and precision farming [8, 9]. This innovative approach has the potential to revolutionize how we approach internet security, making the digital landscape a safer place for all users.

## II. LITERATURE REVIEW

The field of machine learning-driven fraud detection and risk scoring has seen significant advancements in recent years, particularly in the context of credit card fraud and online transactions. This literature review examines key studies and approaches relevant to our proposed platform and browser extension for real-time risk scoring and fraud detection.

Credit card fraud detection has emerged as a primary focus for researchers due to the prevalent use of credit cards in online transactions. [10] proposed an enhanced credit card security system using machine learning for fraud detection. Their work highlights the importance of real-time analysis in identifying fraudulent activities, which aligns with our platform's objectives. The study demonstrates that machine learning models can achieve high accuracy rates in detecting credit card fraud, with some algorithms reporting success rates exceeding 95% in identifying fraudulent transactions. In a related study, [11] explored the application of advanced machine learning techniques in enhancing fraud detection within the banking industry. They evaluated the performance of various models, including LightGBM, XGBoost, CatBoost, vote classifiers, and neural networks, on a comprehensive dataset of banking transactions. The CatBoost model exhibited the highest accuracy in identifying fraudulent instances, showcasing its superior performance. This research underscores the potential of advanced machine learning approaches in mitigating financial losses and ensuring secure transactions, ultimately bolstering trust and security in the banking sector.

One significant challenge in fraud detection is dealing with imbalanced datasets, where legitimate transactions vastly outnumber fraudulent ones. This imbalance can lead to biased models that fail to accurately identify fraudulent activities. [12] investigated this issue in their study on credit card payment fraud detection. They compared the efficacy of two approaches: random under-sampling and oversampling using the synthetic minority over-sampling technique (SMOTE). Their findings revealed that random under-sampling achieves high recall (92.86%) at the expense of low precision, whereas SMOTE achieves a higher accuracy (86.75%) and a more balanced F1 score (73.47%) with a slightly lower recall. This research highlights the importance of addressing class imbalances for effective fraud detection in cybersecurity.

The application of advanced machine learning techniques has proven beneficial in enhancing fraud detection accuracy. [13] explored the use of Light Gradient Boosting Machine (LightGBM) for credit card fraud detection. Their approach involved meticulous preprocessing of anonymized credit card transaction data, emphasizing normalization, feature selection, and strategic data splitting to address class imbalance. The LightGBM model, optimized through Randomized Search Cross-Validation, demonstrated high precision and substantial recall for effective fraud detection, with an impressive Area Under the Curve score. In another study, [14] investigated fraud detection in NoSQL database systems using advanced machine learning techniques. They examined different algorithms, including neural networks, support vector machines, random forests, and clustering, to analyze large database activity logs for identifying anomalous patterns indicative of malicious behavior. Their results showed high accuracy in detecting injection attacks, unauthorized queries, and abnormal database traffic, with low false-positive rates.

As our platform aims to provide real-time risk scoring, recent research on graph-based machine learning techniques is particularly relevant. [15] pioneered a methodology that combines bipartite graph visualization with advanced machine learning techniques for credit card fraud detection. This approach offers a paradigm shift in analyzing and understanding credit card fraud detection systems by integrating machine learning algorithms with network analysis. The use of bipartite graphs serves as a dynamic visual bridge between model predictions and real-world outcomes, enhancing interpretability and facilitating a deeper understanding of the classifier's performance.

The literature review reveals a growing trend towards more sophisticated machine learning techniques for fraud detection and risk scoring. Key themes include addressing imbalanced datasets, employing advanced algorithms for improved accuracy, and leveraging real-time analysis to enhance consumer protection. These findings will inform the development of our machine learning-driven platform and browser extension, ensuring that it incorporates the most effective techniques for real-time risk scoring and fraud detection in website legitimacy verification and consumer protection. By building on existing research and integrating innovative technologies, our platform aims to significantly enhance online security for consumers navigating an increasingly complex digital landscape.

## III. FRAUD DETECTION ML MODELS

In the machine learning-driven platform and browser extension, real-time risk scoring and website legitimacy verification are critical processes powered by multiple machine learning models. These models work in parallel to analyze data in real-time, providing a dynamic assessment of a website's legitimacy. The platform incorporates decision trees, gradient boosting machines (GBM), neural networks, random forests, and support vector machines (SVMs), which collectively form a robust, adaptive system for fraud detection. Each model plays a distinct role in evaluating website features, user behavior, and interaction patterns to flag fraudulent sites.

Decision Trees form the backbone of the system's initial fraud detection process. They are rule-based models that create a series of binary splits in the data based on features such as domain age, SSL certificate validity, and content structure. In real-time, as a user visits a website, the decision tree model processes these features and classifies the website into either a low-risk or high-risk category. For instance, if the domain is less than six months old and lacks a valid SSL certificate, the tree assigns a higher risk score. Decision trees offer an initial classification because they are computationally efficient and can quickly process features to generate an immediate risk score [16]. This provides a first layer of defense by assessing obvious signs of potential fraud.

Random Forests, an ensemble learning technique, expand on the decision tree approach by generating multiple decision trees during training. Each tree in the forest is trained on a random subset of features and data points. In the context of fraud detection, random forests help reduce the risk of overfitting by averaging the results of multiple decision trees, thus improving accuracy and robustness [16]. As a user interacts with a website, the random forest model evaluates features such as the presence of suspicious URLs, abnormal redirections, and suspicious JavaScript activity. By averaging the risk scores generated by each tree, the random forest model provides a more reliable classification, reducing false positives. The diversity of decision trees within the random forest makes it well-suited for identifying a wide range of fraud patterns [17].

Gradient Boosting Machines (GBM) further refine the platform's real-time risk scoring by focusing on iterative improvement. Unlike random forests, which build independent trees, GBMs build each tree sequentially, correcting the errors of the previous tree. This iterative process allows the model to focus on misclassified instances, such as websites that exhibit sophisticated fraud tactics [18]. When a user visits a website, the GBM evaluates complex patterns, such as subtle content changes, phishing attempts disguised as legitimate content, or hidden forms designed to steal personal information. For example, GBMs can give higher weights to behaviors like multiple rapid redirects, which are common in phishing attacks. This process continues as more user interaction data is collected, allowing the model to dynamically adjust the risk score in real-time as it uncovers new fraudulent behaviors [19].

Neural Networks, with their ability to model non-linear relationships and high-dimensional data, add another layer of sophistication to fraud detection. These models are capable of detecting hidden patterns in website features and user behavior that simpler models might miss [17]. For example, a neural network can analyze the structure of a webpage, including metadata, CSS patterns, embedded links, and user interaction data such as mouse movement patterns or the sequence of clicks. The hidden layers of the neural network allow it to detect intricate relationships between these features, such as inconsistencies in a webpage's layout that might indicate a phishing attempt. Neural networks continuously update the risk score based on new data, such as user activity or network traffic, providing an adaptive approach to fraud detection [20].

Support Vector Machines (SVMs) are another model used in the platform for classification, particularly when dealing with highly imbalanced data. SVMs are effective at creating a hyperplane that separates legitimate websites from fraudulent ones based on the input features [16, 21]. In the context of real-time fraud detection, SVMs are trained on features such as the visual appearance of the website (e.g., font and color consistency, image properties) and user behavior data (e.g., clickstream data, time spent on specific elements of the page). By mapping these features into a high-dimensional space, the SVM creates a decision boundary that separates legitimate websites from suspicious ones [22].

To further enhance real-time risk scoring and anomaly detection, the platform also incorporates Autoencoders. These are specialized neural networks designed for unsupervised learning tasks, particularly anomaly detection [23]. Autoencoders work by compressing the input data (features of the website) into a lower-dimensional representation and then reconstructing it. In fraud detection, the autoencoder is trained on legitimate websites, learning to reconstruct typical patterns of website behavior and structure. When the model encounters a fraudulent website, the reconstruction error will be higher because the website deviates from normal behavior [23].

The browser extension integrates these models to provide a seamless real-time fraud detection system. As soon as a user navigates to a website, the extension collects various features in the background, such as website metadata, HTML structure, and user interaction data. These features are fed into the machine learning models—starting with decision trees and random forests for quick initial assessments. The GBM and neural network models then analyze deeper patterns in the user's interaction with the website [18, 19]. The SVM and autoencoder models continuously refine the risk score based on the site's behavior and structure [22, 23].

Overall, the real-time risk scoring and website legitimacy verification processes in the platform leverage multiple machine learning models to analyze features of websites and user interactions from various angles. This multi-model approach ensures that the system adapts to new fraud tactics and provides ongoing protection for users in real-time, helping them navigate the web safely and avoid engaging with fraudulent websites [17, 19, 20].

## IV. REAL-TIME RISK SCORING SYSTEM

One of the key strengths of the real-time risk scoring system is its ability to continuously monitor website behavior and network traffic throughout the entire browsing session, offering a comprehensive and dynamic layer of protection. From the moment a user accesses a site, the system begins analyzing various aspects of the website, starting with basic features such as domain age, SSL certification, and content structure. However, the true power of the system lies in its ongoing ability to track user interactions, such as clickstream data, form submissions, and hovering behavior, identifying any anomalies that may indicate fraudulent activity [24]. For example, if a user fills out a form that requests unnecessary personal information, or if the system detects abnormal mouse movements that resemble typical behavior seen in phishing attempts, the system flags these actions for further scrutiny. Such real-time feedback is invaluable, especially in cases where fraudulent websites are designed to appear legitimate at first but gradually reveal their malicious intent through deceptive interactions.

In addition to tracking user interactions, the platform closely monitors network traffic patterns, which is often a telltale sign of suspicious activity. By analyzing HTTP requests, SSL handshakes, and other network-level interactions, the system can detect traffic anomalies that deviate from standard patterns [25]. For instance, frequent or unexpected requests to third-party domains, especially those with questionable reputations, are clear red flags. Similarly, hidden redirects, where a user is unknowingly sent to multiple or suspicious websites without their knowledge, are immediately flagged as fraudulent behavior. The system also monitors resource requests made by the website, including scripts, images, and cookies, inspecting them for any signs of malicious intent. If a script behaves unexpectedly, such as attempting to execute unauthorized actions in the background, the system raises the risk score accordingly, signaling the potential for malicious behavior [26].

Furthermore, the system's adaptability ensures that it continuously evolves with the changing landscape of online fraud. As new phishing techniques and fraudulent tactics emerge, the risk-scoring models are updated with fresh data to capture these evolving patterns. This means that the system is not only reactive but also proactively learns from new types of fraud [25]. As more websites are analyzed and more data points are collected, the machine learning models become better at detecting previously unseen tactics. This ongoing learning process ensures that users are protected not only from known threats but also from new and emerging fraud strategies. The platform's ability to dynamically adjust to both user behavior and external network conditions, combined with the continuous integration of new fraud patterns, makes it an essential tool in safeguarding users from the increasingly sophisticated landscape of online threats [24].

## V. BROWSER EXTENSION

The browser extension is envisioned as a critical component of the fraud detection platform, aimed at providing real-time protection for users as they browse websites. Its design integrates seamlessly with popular web browsers and operates in the background without the need for user intervention [27]. Upon visiting a website, the extension begins analyzing key elements of the site, including domain metadata, SSL certificate validity, content structure, and network requests. These data points are fed into the platform's machine learning models, which generate an initial risk score that reflects the website's legitimacy [4]. The extension serves as the primary interface for this system, delivering real-time feedback in the form of notifications or warnings when a website exhibits potentially fraudulent behavior.

One of the most important aspects of the extension is its ability to offer continuous monitoring during the user's session on a website. Beyond just analyzing the initial load of the site, the extension tracks ongoing interactions, such as clickstream data, form submissions, and hovering patterns [28]. These interactions are crucial because fraudulent websites often appear legitimate on the surface but reveal their malicious intent through deeper engagement, such as attempting to capture sensitive information via forms or executing hidden scripts. For example, if a user begins to interact with suspicious elements on a page, like a login form that leads to an unknown domain or an unverified external resource, the browser extension would update the risk score in real-time and issue a warning [29]. The extension also monitors for hidden actions, such as redirects to external domains or unauthorized requests to third-party servers, which are common techniques used by phishing and fraudulent websites to mislead users [28].

In addition to its analytical capabilities, the user interface (UI) of the browser extension is designed to be intuitive and informative. Users are notified of potential risks through clear, color-coded alerts. For example, a green indicator might signal that the website is safe, while a yellow or red warning could alert the user to moderate or high risks [30]. The extension would also allow users to click for more detailed information about why a particular site was flagged, such as whether it lacks a valid SSL certificate, redirects to suspicious domains, or exhibits unusual loading behavior [31]. This functionality would empower users to make informed decisions about whether to proceed or avoid engaging with the site. The extension would further provide insight into the website's real-time risk score, showing how different features contributed to the overall risk assessment, enhancing user trust in the system [4].

The browser extension also emphasizes efficiency and low resource consumption, ensuring it does not negatively impact the user's browsing experience. By offloading the more computationally heavy processes to the backend servers where the machine learning models are hosted, the extension maintains lightweight operation on the client side [32]. Privacy is another key focus, with all user data protected through encryption and secure communication protocols. Sensitive information, such as login credentials or personal data, would not be stored or shared, ensuring that while users are protected from fraudulent websites, their own privacy is safeguarded [33]. Overall, the browser extension is a crucial real-time defense mechanism, continuously monitoring, evaluating, and protecting users from fraudulent websites while maintaining a seamless and secure browsing experience [28].

## VI. BROWSER EXTENSION WORKFLOW

The diagram below illustrates the operational workflow of a browser extension designed to enhance online security by detecting potential threats and verifying the legitimacy of websites using machine learning (ML) algorithms. The process initiates when a user navigates to a website, activating the browser extension. This extension plays a crucial role in the security framework by monitoring and analyzing the URLs that users visit in real-time.

As the website loads, the extension captures the URL and sends it to a backend system where it is cross-referenced against a pre-existing database. This database contains a list of URLs that have been previously analyzed and classified as either legitimate or potentially harmful. The initial check against this database serves as the first line of defense; if the URL is recognized and classified as safe, the website is deemed secure for browsing, and no further action is taken. Conversely, if the URL is not found in the database or requires further scrutiny, it undergoes a more detailed examination.

In cases where the URL's legitimacy is uncertain, the system employs machine learning algorithms to analyze the URL and the content of the website. These algorithms assess various characteristics of the website, such as the structure, the nature of embedded scripts, and the overall behavior of the site. By leveraging patterns learned from vast datasets of both fraudulent and legitimate websites, the ML algorithms can effectively determine whether a new URL exhibits typical characteristics of known threats.

The outcome of this analysis leads to one of two possible actions:

• If the ML algorithms classify the website as legitimate, the process concludes, confirming that the website is safe for the user to browse. This decision is communicated back to the browser extension, which allows the user to continue their activity without interruption.

• If the website is determined to be potentially malicious, the system triggers a threat alert. This alert is sent back to the extension, which then notifies the user of the potential risk. The alert may advise caution, offer detailed information about the nature of the detected threats, or even block access to the site to prevent potential harm, depending on the severity of the risk and the configuration of the system.

This proactive approach, integrating both a reactive database check and a predictive ML analysis, ensures that users are protected from both known threats and new, emerging risks. The browser extension thus acts not only as a shield against potential online threats but also as a dynamic tool that adapts to the evolving landscape of internet security, continually updating its algorithms and database to incorporate the latest data and trends in cybersecurity.

## VII. BROADER IMPLICATIONS FOR CONSUMER PROTECTION

The development of the browser extension and the underlying fraud detection platform plays a crucial role in advancing consumer protection within the realm of e-commerce. By providing real-time analysis and threat detection, the platform significantly enhances the security posture for online shoppers, contributing to a safer e-commerce environment. The implementation of machine learning algorithms enables the system to identify and react to potential threats swiftly, ensuring that consumers are less likely to fall victim to fraud and scams. This protection is vital as e-commerce continues to grow, with more consumers relying on online platforms for their shopping needs [7]. The ability to securely navigate websites and avoid deceptive or malicious sites helps in maintaining consumer confidence in digital marketplaces, which is essential for the sustained growth of the e-commerce sector.

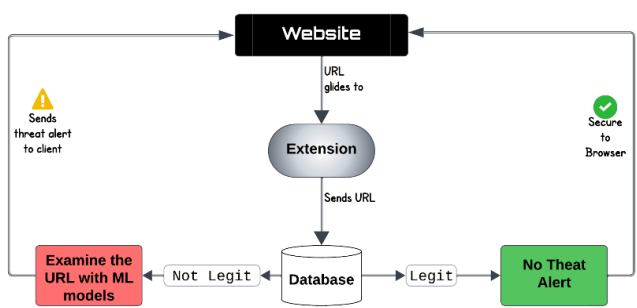

Furthermore, the platform's impact extends beyond individual consumer protection to national security. Online fraud schemes often target not only individuals but also the institutions and infrastructures that underpin the national economy. By preventing large-scale fraud attacks, the system helps safeguard critical infrastructures that could be undermined by sophisticated cybercriminal activities. For instance, preventing fraud in financial transactions and protecting users from phishing sites help maintain the integrity of financial institutions [31]. Additionally, by securing consumer data against breaches, the platform indirectly supports broader national efforts to combat cybercrime. This is increasingly important as cyber threats become more complex and pervasive, potentially threatening economic stability and security.

Moreover, the incorporation of data visualization techniques into the platform enhances the understanding and accessibility of financial information, making it easier for users to recognize patterns and anomalies in their financial transactions. This aspect of data visualization in finance is crucial as it helps bridge the gap between complex data analyses and user-friendly interfaces, empowering consumers to make informed financial decisions and detect potential fraud independently [34]. The role of machine learning in this context is also pivotal for policy making, as it provides policymakers with actionable insights derived from data analysis. These insights can inform regulations and policies that protect consumers and the economy from the adverse effects of online fraud [35, 36].

Lastly, the browser extension and platform serve as a deterrent against the proliferation of online fraud tactics. By continuously updating its threat detection models and adapting to new fraud strategies, the platform ensures that emerging threats are swiftly identified and mitigated. This ongoing adaptation not only protects individual consumers but also contributes to a broader cybersecurity ecosystem that is resilient against evolving threats. The collective data and insights gained from the platform's operations can also inform policy-making and regulatory measures, enhancing overall cyber defense strategies at a national level [37]. In essence, the platform not only defends against immediate threats to consumers but also contributes to the long-term resilience and security of national economic and infrastructural systems, reinforcing the importance of integrated cybersecurity measures in today's digital age [38].

## VIII. CONCLUSION

This paper has detailed the design and functionality of a machine learning-driven platform and browser extension aimed at enhancing online security through real-time fraud detection and website legitimacy verification. The platform integrates various machine learning models, including decision trees, gradient boosting machines, neural networks, random forests, and support vector machines, each of which contributes to analyzing website features, user interactions, and network traffic [39, 40]. Together, these models enable the system to assess and respond dynamically to potential threats. The browser extension serves as the primary interface for users, providing real-time notifications and warnings about potentially fraudulent websites, thereby empowering users to avoid harmful interactions.

The framework presented in this paper offers valuable insights that could inform the development of future online security systems. By demonstrating the effectiveness of combining multiple machine learning models to assess and mitigate online threats in real-time, this approach highlights the potential for such systems to evolve with the ever-changing nature of cybersecurity risks. The system's ability to process real-time data and deliver user-focused feedback ensures that security mechanisms are both proactive and adaptable. These principles could greatly influence the development of future fraud detection and website legitimacy verification systems, helping them to balance the need for robust security with user accessibility and efficiency [41]. Additionally, as demonstrated in research on gender diversity in top management teams, diverse perspectives play a key role in improving decision-making outcomes and risk management [42, 43]. This insight aligns with the use of diverse data sources and machine learning models in fraud detection, which enhances the platform's capacity to identify and respond to sophisticated and evolving fraud tactics [44].

Looking ahead, several areas for future development could enhance the platform's scalability and effectiveness. One key avenue is scaling the system to handle larger data sets and higher traffic volumes while maintaining performance efficiency. This could involve optimizing current machine learning models or exploring new algorithms capable of faster processing times [45]. Furthermore, integrating advanced techniques such as **reinforcement learning** could enable the platform to learn and adapt more effectively from each interaction, thus improving its ability to detect emerging fraud patterns with greater accuracy [17].

Another promising direction involves expanding the platform's predictive capabilities. Using advanced analytics to not only detect but also predict future threats based on trending data and user behavior could revolutionize the approach to fraud detection, offering protection before users are exposed to risks [19]. These enhancements will allow the platform to evolve alongside the digital landscape, ensuring that it remains effective in combating increasingly complex cybersecurity threats [22].

Finally, the platform and browser extension described in this paper provide a comprehensive solution to the challenges of online fraud detection. By leveraging multiple machine learning models and continuously adapting to new threats, the system offers significant potential to improve consumer protection in the digital age. As online interactions continue to grow in importance, the need for advanced, adaptive security systems like the one proposed here will become increasingly critical in safeguarding users' data and maintaining trust in digital platforms.